\documentclass[11pt]{article}
\usepackage{blois,epsfig}

\def\Journal#1#2#3#4{{#1} {\bf #2}, #3 (#4)}

\def\NCA{\em Nuovo Cimento}

\def\PRD{{\em Phys. Rev.} D}

\def\APJ{\em Astrophys. J.}
\def\GRG{\em Gen. Rel. Grav.}
\def\NAT{\em Nature}

\def\be{\begin{equation}}
\def\ee{\end{equation}}
\def\bea{\begin{eqnarray}}
\def\eea{\end{eqnarray}}

\begin{document}

\vspace*{4cm}

\title{IS THE CARTER-ISRAEL CONJECTURE CORRECT?}

\author{COSIMO BAMBI}

\address{Institute for the Physics and Mathematics of the Universe,
The University of Tokyo, \\ Kashiwa, Chiba 277-8568, Japan}

\author{KATHERINE FREESE}

\address{The Michigan Center for Theoretical Physics,
Department of Physics, University of Michigan, \\ 
Ann Arbor, Michigan 48109, USA}

\author{ROHTA TAKAHASHI}

\address{Cosmic Radiation Laboratory, 
Institute of Physical and Chemical Research, \\ 
Wako, Saitama 351-0198, Japan}

\maketitle

\abstracts{According to the Carter-Israel conjecture, 
the end-state of the gravitational collapse of matter 
is a Kerr-Newman black hole. Nevertheless, neither the 
theory nor observations can confirm that. In this talk, 
we discuss the possibility that the collapsing matter 
can create a super-spinning compact object with no event 
horizon, and we show how near future observations at 
sub-millimeter wavelength of SgrA* can test this 
scenario for the black hole candidate in the Galactic 
Center.}

\section{Introduction}

The question about the nature of the end-state of the 
gravitational collapse of matter is a long-standing
issue which is still unsolved. At the theoretical
level, the question might be addressed by general
relativity (GR), our current theory of gravity. Here 
we know some singularity theorems which show that
collapsing matter leads to the formation of
singularities. Basically there are two possibilities: 
if the singularity is hidden behind an event horizon, 
the final product is a black hole (BH), if it is
not, we get a naked singularity (NS). However, the 
true question is if ``real matter'' can form a NS. 
That is unknown, even because it is not so easy to say 
what ``real matter'' must be. In addition to this, it 
is not obvious that GR can address the issue: the 
theory has been tested only in the weak field limit 
and we do not know if it works in the case of strong 
gravity.

As for observations, today we have clear evidences 
supporting the existence of super-massive bodies at the 
center of many galaxies and compact stellar-mass bodies 
in the Galaxy. These objects are very likely the final 
product of the gravitational collapse of matter and are
believed to be BHs, but actually there are no evidences 
that they have an event horizon and we do not know if 
the spacetime around them is like the one predicted by GR.

Even if GR allows for the creation of NSs, their existence
seems to be problematic: in a spacetime containing a NS, it 
is typically possible to go back in time and therefore to 
violate causality. So, it is common opinion that NSs cannot 
be created by any physical process (Cosmic Censorship 
conjecture)~\cite{penrose}. When we consider the case of 
collapsing matter, this idea leads to the Carter-Israel 
conjecture: the final product of the gravitational collapse 
of matter is a Kerr-Newman BH. The latter is an object 
characterized by just three parameter; that is, the mass 
$M$, the electric charge $Q$, and the spin $J$ (or the Kerr 
parameter $a=J/M$). These three parameters are not completely 
free, but must satisfy the relation $M^2 > Q^2 + a^2$, 
which is just the condition for the existence of the horizon.
In what follows, we can restrict our discussion to the case
$Q = 0$, because the electric charge is usually negligible
for large astrophysical bodies.

\section{Motivations for new physics}

It is well known that simple considerations suggest that
the Planck scale, $E_{Pl} \sim 10^{19}$~GeV, is the natural 
UV cut-off of classical GR. In this case, the theory would
be unable to describe phenomena in which the characteristic
energy exceeds $E_{Pl}$. If we apply this idea to the case
of a Kerr spacetime with $M < |a|$, where observer-independent 
quantities like the scalar curvature diverge at the singularity, 
it is at least questionable to expect that the GR prediction
of the violation of causality is reliable: the latter requires 
that a particle coming from infinity ``enters'' into the 
singularity~\cite{chandra}! On the other hand, new physics may 
replace the singularity with something else and Nature may 
conserve causality not because it is impossible to create a NS, 
but because there is no singularity in the full theory. Adopting 
this point of view, there are no fundamental reasons that forbid
the creation of a super-compact object with $M < |a|$ in the
Universe.

\section{Direct image of the accretion flow of SgrA*}

In this talk, we discuss the possibility of testing the
Carter-Israel conjecture by observing the direct image of the
accretion flow onto the BH candidate in the Galactic Center.
This is a short summary of the material presented in 
Bambi and Freese~\cite{bambi} and Takahashi 
{\it et al.}~\cite{takahashi}.

\subsection{Present observations}

Because of the strong gravitational field, the light
passing near compact objects does not go along straight
lines, but bends. The result is that the apparent size
of a compact object seen by a distant observer is 
always larger than the real size of the object.

In Doeleman {\it et al.}~\cite{doeleman}, the authors 
reported the observation at the wavelength of 1.3~mm of 
the radio source SgrA*, which is coincident with the 
position of the BH candidate in the Galactic Center at 
the level of 10~mas. Modeling SgrA* as a circular Gaussian 
brightness distribution, they find that the intrinsic
diameter of the radio source is $37^{+16}_{-10}$~$\mu$as
at 3$\sigma$. Nevertheless, if SgrA* were a spherically
symmetric photosphere centered on the BH, one would expect
that the minimum apparent diameter would range from
52~$\mu$as (Schwarzschild BH) to 45~$\mu$as (Kerr BH with 
$|a| = M$ for an observer on the equatorial plane). Although 
the current data are not yet
capable of absolute confirmation of such a measurement,
its implications could be intriguing. One possibility is
that the radio source is not perfectly centered on the 
BH~\cite{doeleman}. A second possibility is that SgrA*
is centered on the compact object, but the latter is a 
super-rotating object with $|a| > M$~\cite{bambi}.
In the latter case, the emission region could have a smaller 
apparent size: since the compact object has no event horizon,
the actual size of the photosphere could be very small. 
That is impossible for the case of BH, just because the 
photosphere has to be out of the event horizon.

\subsection{Future observations}

It is common opinion that, for frequencies larger than about
500~GHz (wavelengths smaller than 0.6~mm), the plasma around
the BH candidate in the Galactic Center becomes optically
thin. At such frequencies, one should observe the ``shadow'' 
of the BH~\cite{melia}, a dark area over a bright background.

\begin{figure}[p]
\begin{center}
\psfig{figure=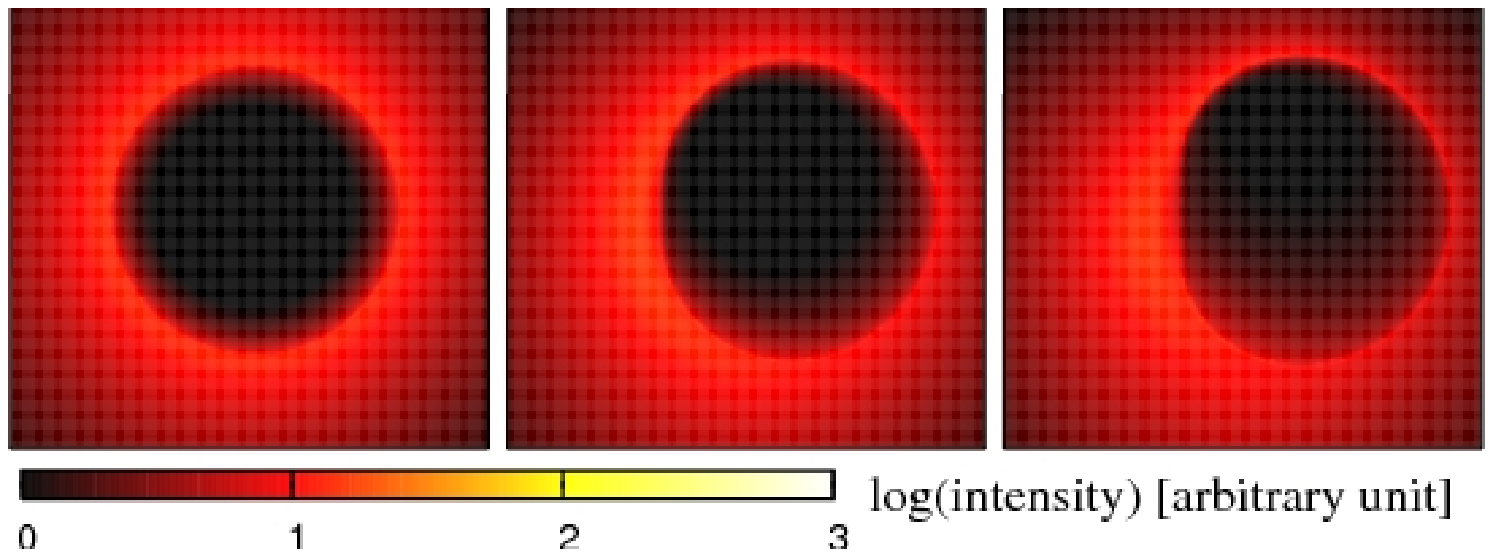,width=150mm} \\ \vspace{-0.030cm}
\psfig{figure=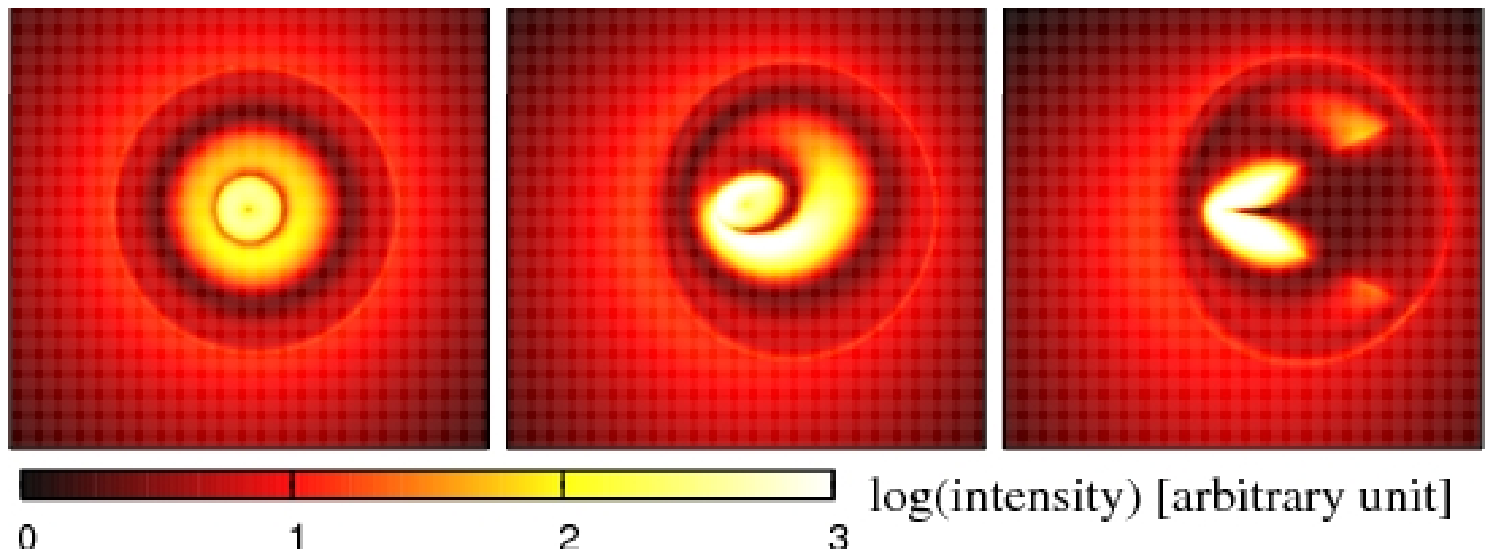,width=150mm} \\ \vspace{-0.030cm}
\psfig{figure=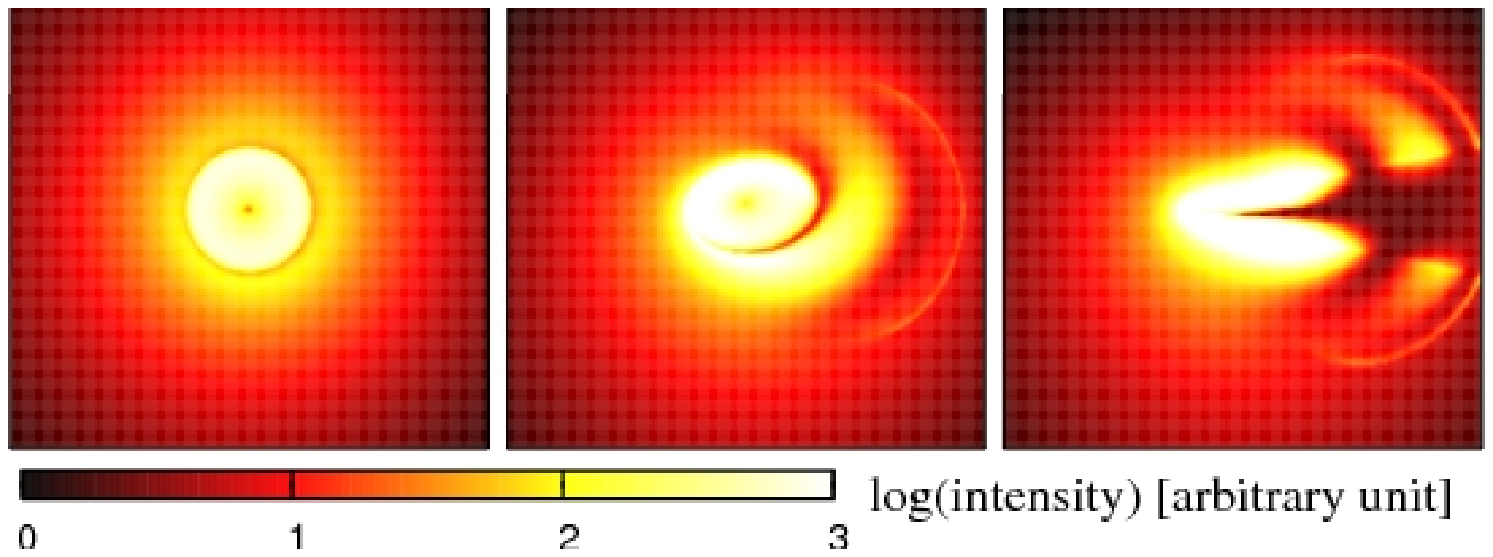,width=150mm}
\end{center}
\caption{\label{fig}
Images of the accretion flow around a black hole with 
$a/M = 0.9999$ (top panels) and super-spinning black objects
with $a/M = 1.0001$ (central panels) and $a/M = 2$ (bottom panels).
The viewing angle is $i = 5^\circ$ (left column), $i = 45^\circ$ 
(central column), and $i = 85^\circ$ (right column).}
\end{figure}

Assuming that the particles of the accreting gas follow
marginally bound time-like orbits of the background metric,
have zero component of the angular momentum along the BH
spin at infinity, and an emissivity function independent of 
frequency and proportional to $1/r^2$, the image of the 
accretion flow onto a BH with $a/M = 0.9999$ would look like 
the pictures in the top panels of Fig.~\ref{fig}. Very long 
baseline interferometry observations are expected to be able 
to see the shadow of the BH candidate in the Galactic Center 
in a few of years.

However, if the super-massive object in the Galactic Center
is not a BH, the image of its accretion flow at sub-millimeter
wavelengths would be different. Fig.~\ref{fig} shows the
image of a super-spinning object with $a/M = 1.0001$ (central
panels) and $a/M = 2$ (bottom panels), assuming that the NS is
replaced by a spherical object with radius $0.1 \, M$ and
a perfectly absorbing surface~\cite{takahashi}.

The exact prediction of the intensity map of the accretion flow 
around a compact object depends on the features of the accretion 
flow. Nevertheless, there is an important difference between BH 
and super-spinnning object. If the first case, there is always the 
shadow, i.e. the dark region over the bright background. In the 
second case, the absence of the event horizon let the observer
see radiation coming from regions at very small radii: the result
is the presence of some brighter spots inside the shadow. Indeed, 
close to the singularity, the gravitational redshift is negligible 
and one would even expect that the temperature of the plasma is 
higher. Both features imply a higher luminosity in the region where
instead one should expect lower luminosity in the case of BH.

\section*{Acknowledgments}

C.B. was supported by World Premier International Research 
Center Initiative (WPI Initiative), MEXT, Japan.
K.F. thanks the DOE and the MCTP at the University of 
Michigan for support.
R.T. was supported by the Grant-in-Aid for Scientific
Research Fund of the Ministry of Education, Culture, Sports, 
Science and Technology, Japan [Young Scientists (B) 21740149].

\end{document}